\documentclass{JINST}

\title{Signal Processing and Electronic Noise in LZ}

\author{D. Khaitan$^a \thanks{Corresponding author.}$ for the LZ Collaboration\\
\llap{$^a$}Department of Physics and Astronomy, University of Rochester\\
Rochester, NY - 14627, USA\\
E-mail: \email{dkhaitan@u.rochester.edu}}

\abstract{The electronics of the LUX-ZEPLIN (LZ) experiment, the 10-tonne dark matter detector to be installed at the Sanford Underground Research Facility (SURF), consists of low-noise dual-gain amplifiers and a 100-MHz, 14-bit data acquisition system for the TPC PMTs. Pre-prototypes of the analog amplifiers and the 32-channel digitizers were tested extensively with simulated pulses that are similar to the prompt scintillation light and the electroluminescence signals expected in LZ. These studies are used to characterize the noise and to measure the linearity of the system.  By increasing the amplitude of the test signals, the effect of saturating the amplifier and the digitizers was studied. The RMS ADC noise of the digitizer channels was measured to be 1.19 $\pm$ 0.01 ADCC.  When a high-energy channel of the amplifier is connected to the digitizer, the measured noise remained virtually unchanged, while the noise added by a low-energy channel was estimated to be 0.38 $\pm$ 0.02 ADCC (46 $\pm$ 2$\mu$V). A test facility is under construction to study saturation, mitigate noise and measure the performance of the LZ electronics and data acquisition chain.} 


\begin{document}

\section{Introduction}

The LUX-ZEPLIN (LZ) experiment is a second-generation dark-matter detector to be installed in the Davis Cavern at the Sanford Underground Research Facility (SURF) \cite{LZCDR}. LZ contains 10 tonnes of liquid xenon with a 7-tonne active volume in a two-phase time projection chamber (TPC). The detector will be housed in the same water tank that is currently used for the Large Underground Xenon (LUX) dark-matter detector to shield it from environmental radiation \cite{LUX}. To further reduce the background level in the detector, the TPC is surrounded by a  gadolinium-loaded liquid scintillator active veto. The scintillator and a portion of the water tank are viewed by 120 PMTs. Another 180 PMTs, mounted inside the cryovessel, monitor the thin layer of xenon between the cryostat and walls of the TPC. 

\begin{figure}
    \centering
    \includegraphics[width=6in]{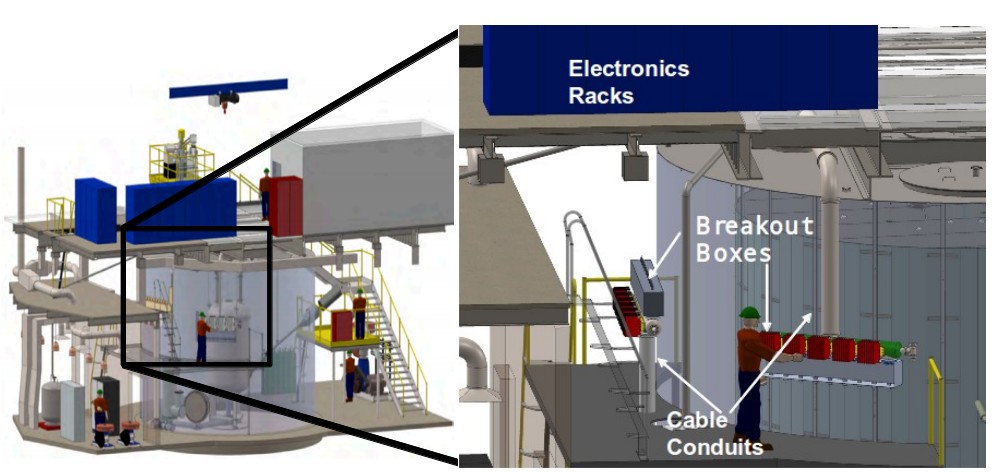}
    \caption{\textbf{\textit{(Left)}} Overview of LZ installed in the Davis Cavern. \textbf{\textit{(Right)}} Close-up of the mezzanine level showing the breakout boxes and the cable conduits. The electronics racks on the upper level are also indicated.}
    \label{fig:overview}
\end{figure} 

The TPC employs 488 Hamamtsu R11410 3-inch-diameter PMTs \cite{LZCDR}, distributed in two arrays, one on the top with 247 PMTs in a hexagonal configuration with two circular outer rows and one on the bottom with 241 PMTs distributed in a close-packed hexagonal configuration. The high voltage (HV) and signal cables for these PMTs are routed to a mezzanine level where two breakout boxes are installed, as shown in Fig. ~\ref{fig:overview}. In this paper we describe tests carried out on pre-prototype components of this electronics chain and discuss a setup that is being assembled to test the full electronics and data acquisition chain.

\section{Preliminary Tests of the LZ TPC Electronics Chain}

The HVs for TPC PMTs are provided by Weiner MPOD EDS 20130x\_504 modules \cite{Wiener}. Kerpen cables with 48-cores feed HV to the flanges on the breakout box from the electronics rack, each of which provide HV for 32 PMTs \cite{Kerpen}. Gore 3007 cables are used to supply HV from the breakout box to the base of the PMTs and also to transport the signals from the PMT base to the breakout box \cite{Gore}. These cables have been chosen for their low heat load and low signal attenuation. The area attentuation was measured with 45 and 90 feet of these cables with 2-ns risetime pulses and found to be 18\% and 28\%, respectively. The amplitude attentuation was found to be 56\% and 80\%, respectively. 

The signals of the TPC PMTs are amplified and shaped with dual-gain amplifiers. The amplifier cards are located in cages on the breakout boxes with each cage containing four amplifier cards with eight input channels each. The high-energy (HE) output of the pre-prototype amplifier card has a full-width at tenth maximum (FWTM) shaping time of 30 ns and an area gain of 0.5, while the low-energy (LE) output has a FWTM shaping time of 60 ns and an area gain of 20. The total front end LZ detector electronics noise requirement for the amplifiers and digitizer is <0.5 mV. 

\begin{figure}
    \centering
    \includegraphics[width=5in]{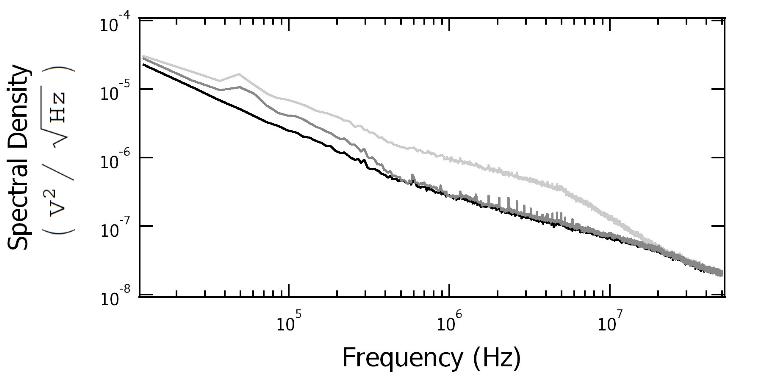}
    \caption{Averaged noise spectrum for the pre-prototype amplifer and digitizer chain. The trace of the noise spectrum for the DDC-32 is shown in black. The noise spectra for a combinations of the DDC-32 and the LE and HE channels is shown in light and dark grey, respectively.}
    \label{fig:noise}
\end{figure} 

The output signals from the amplifiers are routed to electronic racks located directly above the breakout boxes on the upper level of the experimental hall, using LMR-100-FR cables \cite{LMR}. These cables are low smoke and zero halogen (LSZH) cables and have a low signal area attenuation.

The output signals of the amplifier are digitized using DDC-32 digitizers built by Skutek  Instrumentation \cite{DDC32}. The digitizers have 32 input channels, 2V dynamic range, 14 bit resolution and a sampling frequency of 100 MHz. The RMS noise of the DDC-32 is 1.19 $\pm$ 0.02 ADCC, increasing to 1.22 $\pm$ 0.02 ADCC when a HE output channel of the amplifer is connected to it and 1.58 $\pm$ 0.02 ADCC in the case of the LE channel. The HE channel adds almosts no noise to the inherent noise of the digitizer and the LE channel of the amplifier adds 1.04 $\pm$ 0.02 ADCC of noise in quadrature.

The amplifier and digitizer chain was further studied to determine if the RMS noise was dominated by noise at a particular frequency. For each of the three combinations (DDC-32, HE+DDC-32 and LE+DDC-32), a thousand waveforms of 8,192 samples each were collected. The magnitude squared of the Fourier transforms were averaged together and these spectra are shown in Fig. ~\ref{fig:noise}. All three spectra fall as 1/$\sqrt{Hz}$ and there are no dominant noise contributing frequencies. As was expected, the HE channel adds almost no noise to the noise of the digitizer. The LE channel adds noise uniformly over a broad range of frequencies. 

\begin{figure}
    \centering
    \includegraphics[width=6in]{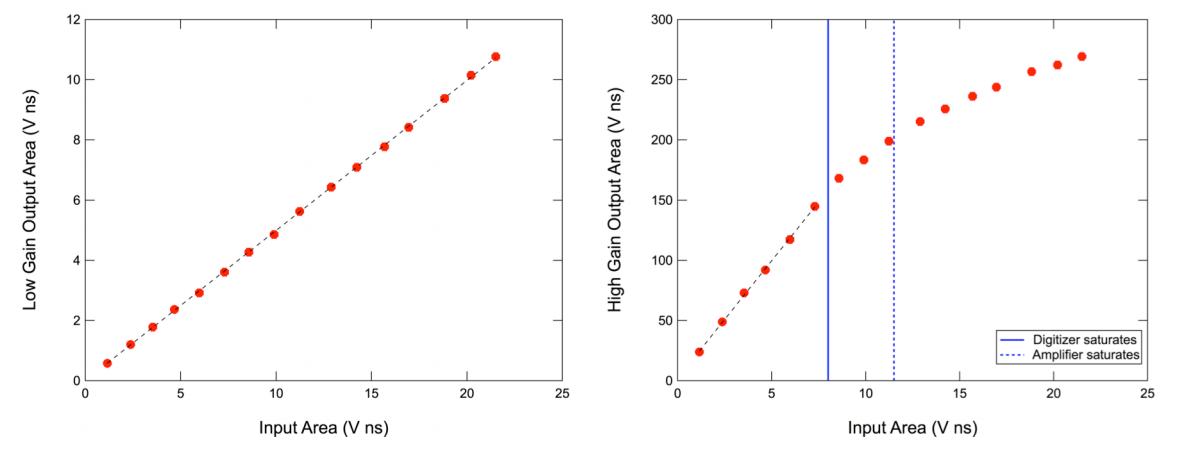}
    \caption{Gain linearity plots generated with the pre-prototype amplifier. The pre-prototype amplifiers have an area gain factor of 0.5 for the HE channel (left) and an area gain factor of 20 for the LE channel (right).}
    \label{fig:linearity}
\end{figure} 

The response of the pre-prototype amplifier and digitizer were tested for area linearity with two types of pulses: a fast pulse with 10\%-90\% rise times in the range 10-50 ns and fall times two and a half times the rise time and with a slower pulse of Gaussian shape with rise times in the range of 150-1500 ns. The input pulse amplitude to the amplifier was varied between 0.020 V and 0.50 V. For the results shown in Fig.~\ref{fig:linearity}, a fast pulse with a rise time of 20 ns was used. A thousand waveforms were averaged to produce each point and the standard deviation of the population is <1\%. For all tested pulses, the measured area gains match specifications. For slower pulses the saturation points moved to larger input areas, thereby giving us a pulse width dependent dynamic range for LZ \cite{Jun}.

\section{Full Electronics Test Facility}

\begin{figure}
    \centering
    \includegraphics[width=6in]{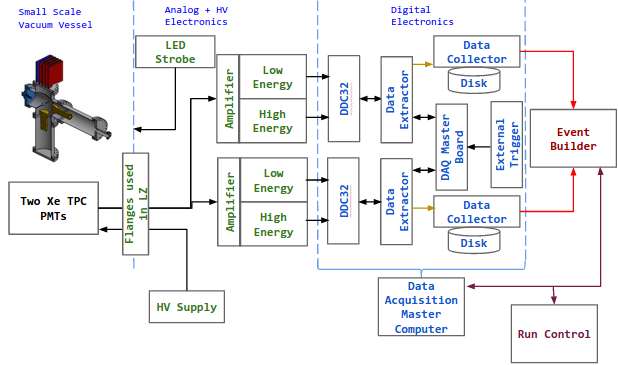}
    \caption{Schematic overview of the setup for the electronics chain test. The direction of signal flow is indicated as are the various sections of the setup. }
    \label{fig:chaintest}
\end{figure} 

To test all the components of the LZ data acquisiton system, a electronics chain test facility is being created at the University of Rochester. This facility will include all the LZ components from PMT to disk, as shown in Fig.~\ref{fig:chaintest}. The purpose of this facility is to provide a development system to extensively test and measure the performance of hardware, software, and firmware, before deployement in LZ. Having all the components from PMT to digitizer will provide the opportunity to study the noise in the full electronics chain. 

This setup will include two Hamamatsu R11410 PMTs in a small vacuum vessel, which includes a feed through for an LED signal. The same signal and HV flanges to be used on the breakout boxes are used on the vessel. Only one feed-through connector on the HV flange will be populated to bias the two PMTs. The signal flange is mounted with an amplifier cage and two amplifier cards. These extra channels from the amplifier will be digitized to measure crosstalk and develop methods to mitigate common mode noise.

The signals from the amplifiers will be digitizied with two DDC-32s. Only when a valid event is identified from the external trigger in the DAQ Master is data offloaded from the DDC-32 to disk on the Data Collector via the Data Extractor \cite{Eryk}. This entire digital chain will be monitored and controlled by the Data Acquisition Master computer.

The Event Builder reads data from the disks on the Data Collectors and builds event files. Run Control will initiate the start and stop of the data acquisiton, as well as monitor its status and the status of the Event Builder. Run Control will also be responsible for communicating the run parameters to the Data Acquisition Master Computer and the Event Builder. 

The use of an LED will allow a range of components in the chain to be tested. Small, well timed pulses will be generated in the LED to study relative timing difference between channels, while larges pulses will be used to study saturation in the PMTs, the amplifiers, and the digitizers. Using a high pulse rate provides the opportunity to stress test the data throughput and benchmark the performance of the acquisition system. Currently, the vacuum vessel is being assembled and firmware and software are being deployed.

\section{Summary}

Testing has been completed on pre-prototypes of the amplifier and digitizer and they have met noise and area linearity requirements. The process of constructing an electronics test facility to mimic the setup of LZ TPC PMTs has begun. Using this setup we will study the effects of saturation as well as possible time difference between channels on the PMT, amplifier, and digitizer and develop techniques to mitigate noise. This setup will also be used to optimize the preformance of the  data acquisiton and event building hardware and software.

\acknowledgments

The work was partially supported by the U.S. Department of Energy (DOE) under award numbers DE- SC0012704, DE-SC0010010, DE-AC02-05CH11231, DE-SC0012161, DE-SC0014223, DE-FG02- 13ER42020, DE-FG02-91ER40674, DE-NA0000979, DE-SC0011702, DE-SC0006572, DE-SC0012034, DE-SC0006605, and DE-FG02-10ER46709; by the U.S. National Science Foundation (NSF) under award numbers NSF PHY-110447, NSF PHY-1506068, NSF PHY-1312561, and NSF PHY-1406943; by the U.K. Science \& Technology Facilities Council under award numbers ST/K006428/1, ST/M003655/1, ST/M003981/1, ST/M003744/1, ST/M003639/1, ST/M003604/1, and ST/M003469/1; and by the Portuguese Foundation for Science and Technology (FCT) under award numbers CERN/FP/123610/2011 and PTDC/FIS-NUC/1525/2014.


\begin{thebibliography}{9}

\bibitem{LZCDR}
D.S. Akerib et al., 
\emph{LUX-ZEPLIN (LZ) Conceptual Design Report}, 
\href{http://arxiv.org/abs/1509.02910}
{[arXiv:physics.ins-det/1509.02910]}.

\bibitem{LUX}
D.S. Akerib et al.,
\emph{The Large Underground Xenon (LUX) experiment}
\href{http://www.sciencedirect.com/science/article/pii/S0168900212014829http://www.sciencedirect.com/science/article/pii/S0168900212014829}{Nucl. Instrum. Meth. A 704 (2013) 111-126.}

\bibitem{Wiener}
Weiner HV System
\href{http://www.wiener-d.com/sc/power-supplies/mpod--lvhv/mpod-hv-module.html}{\emph{http://www.wiener-d.com/sc/power-supplies/mpod--lvhv/mpod-hv-module.html}}

\bibitem{Kerpen}
Kerpen Cable
\href{http://www.leoni-industrial-projects.com/uploads/tx_downloadleoni/en_Power_and_Control__Cables_web_01.pdf}{\emph{http://www.leoni-industrial-projects.com/}}

\bibitem{Gore}
Gore Cable
\href{http://www.gore.com/en_xx/products/cables/index.html}{\emph{http://www.gore.com/}}

\bibitem{LMR}
LMR Cables
\href{http://www.timesmicrowave.com/cms/products/cables/lmr/}{\emph{http://www.timesmicrowave.com/cms/products/cables/lmr/}}

\bibitem{DDC32}
Skutek Instrumentation
\href{http://www.skutek.com/}{\emph{http://www.skutek.com/}}

\bibitem{Jun}
J. Yin
\emph{The Dynamic Range of LZ}
(JINST) {2015}.

\bibitem{Eryk}
E. Druszkiewics,
\emph{The Data Acquisition System for LZ},
{JINST} (2015).

\end{thebibliography}
\end{document}